\documentstyle[11pt,newpasp,twoside,epsf]{article}
\def\Msun{M_\odot}
\def\pc{\, {\rm pc} }
\def\kpc{\, {\rm kpc} }

\def\kms{\, {\rm km \, s}^{-1} }
\def\BF{\mathbf}
\def\bey{\begin{eqnarray}}
\def\eey{\end{eqnarray}}
\def\beq{\begin{equation}}
\def\eeq{\end{equation}}

\def\edcomment#1{\iffalse\marginpar{\raggedright\sl#1\/}\else\relax\fi}
\reversemarginpar

\begin{document}
\title{Are Omega Centauri and G1 like massive globular clusters
the remnant cores of infalling dwarf galaxies? -- Dynamical constraints}
\author{HongSheng Zhao}
\affil{Institute of Astronomy, Cambridge}
\begin{abstract}
We simulate the result of orbital decay for a dwarf satellite galaxy in a
disk galaxy potential.  The peeling-off of the satellite by the strong
tidal field within the disk of the host galaxy quickly reduces the
satellite's mass, and drastically slows down the decaying of the orbit.
The result is generally a cluster on a large halo orbit, like the M54 cluster
of the Sagittarius galaxy in the Milky Way and the massive G1 cluster in 
outer halo of M31.  It is generally challenging for this
mechanism to deliver globular cluster-mass objects to the observed
very close-in orbit of Omega Cen unless the dwarf galaxy was born within
a few scale lengths of the Galactic disk.
\end{abstract}

\section{Introduction}
Omega Centauri is the most massive star cluster of the Milky Way.
Unlike other Galactic globular clusters, its stellar
population is not coeval with significant spread in metallicity, age and
kinematics (see this proceeding).  It has long been proposed that it
is the dense core left from a stripped-off dwarf satellite, which has
fallen inside the Milky Way, and has dynamically decayed to its
present position of about 6 kpc from the Galactic center (Freeman
1993).  The progenitor could have a dark halo, and could be similar to
that of the Sagittarius dwarf galaxy, which is known to have multiple
stellar populations.  Sgr also has a central massive cluster M54, the
2nd most massive cluster in the Milky Way after Omega Cen.  In this
scenarios the metals produced in the progenitor could be recycled due
to a deep potential well of the dark matter halo of the progenitor.  The
progenitor could be somewhere between a dwarf elliptical (such as M32,
NGC205) or a dwarf spheroidal system (such as Draco, Fornax,
Sagittarius).  The big mass of the progenitor helps to shrink the orbit
by dynamical friction.

While very attractive, the above hypothesis can now be constrained by
the recently measured small proper motion of Omega Cen (Dinescu et
al. 1999), which implies that it is on a retrograde
eccentric orbit with a pericenter of 1 kpc, and apocenter of 6 kpc.
It is presently close to its apocenter.  The orbit is much more close-in
than typical halo objects, a point which we will come to later
on.  More interestingly, one can ask some more general questions.
What are the possible outcomes of tidal stripping of a dwarf satellite?  Is
this a general way to account for globular clusters?  How often is a
system like Omega Cen produced?  The answers to these questions
will help us to test the validity of the theory of hierarchical
merger formation of galaxies by comparing with observations globular
clusters of the Milky Way.  In this regard, it is interesting to note
that the disk galaxy M31 also has its share of very massive star
clusters.  For example, G1 (or Mayall II) is about 4-5 times the mass of
Omega Cen, and also shows signs of non-coeval stellar populations
(Rich et al. 1996, Meylan et al. 2001).  It is at a projected distance
of about 40 kpc from the center of M31.  A system (NGC1023-13) almost
identical to G1 is also found in the S0 galaxy NGC1023, at about a
projected distance of 40 kpc from the host galaxy (Larsen 2001).  It
has been argued that giant star clusters like G1 and NGC1023-13 are
the transitional objects between a globular cluster and a dwarf galaxy
(Meylan et al. 2001).  In this regard, it is interesting to ask
whether Omega Cen could be a smaller, more advanced version of G1 and
NGC1023-13, which decayed closer to its host galaxy.

Here we simulate the orbital decay histories of a dwarf satellite.  We
argue that the progenitor of Omega Cen in the Milky Way {\it cannot}
be on the large orbits as the progenitors of G1 and NGC1023-13 in
their host galaxies.  The progenitor of Omega Cen must be either on a
surprisingly small orbit from the start, or has a surprisingly dense
halo, which delays the stripping process and allows dynamical friction to
progress for a Hubble time.  In other words, tidal stripping of dwarf
satellite galaxies typically produces systems like M54, G1 and NGC1023-13.
The progenitor of Omega Cen is not on ``typical'' orbits of 
known halo satellite systems.

\section{Models and Methods}

\subsection{Host galaxy potential model}

We model the host galaxy by a simple singular isothermal (SIS) potential
\beq
\phi(R)=V_{\rm cir}^2 \ln R,
\eeq
where $V_{\rm cir}=200\kms$ appropriate for the Milky Way 
and $250\kms$ appropriate for M31.
We call these two models SIS200 and SIS250 respectively.
More detailed models with disks are discussed later.

\subsection{Satellite model}

We model the progenitor as spherical with a broken power law profile where the
mass increases with radius as
\beq
M_s(r) = 1 \times 10^7 \Msun \left({r \over 200 \pc} \right)^p,
\eeq
where the power $p=1$ or $p=2$, and 
we make G1 the transitional object between a dwarf galaxy
and a globular cluster (Meylan et al. 2001).
Fig.1 shows the mass profiles roughtly reproduce
the mass and size relations of M54, Omega Cen and G1.
The extrapolated mass profiles are somewhere between 
the nucleated dwarf ellipticals (M32 and NGC205) and
the more diffuse dwarf spheroidals (Draco and Fornax).

\subsection{Dynamical friction}

We model the orbital decay of the satellite using 
Chandrasekhar's dynamical friction formula 
\beq
{d{\BF V_s} \over dt} = - {GM_s {\BF V_s} \over \left|{\BF V_s}\right|^3}
\left[4\pi \xi G\rho(R)\right],
\eeq
where 
\beq
\rho(R) = {V^2_{\rm cir} \over 4 \pi G R^2}
\eeq
is the density of the dynamical matter in the host galaxy at radius R for
our SIS model (eq. 1), 
and the $\xi$ is some demensionless function of the speed $|{\BF V_s}|$, 
including the Coulomb logarithm.
Here we assumed an isotropic velocity distribution of the
stars in the host galaxy with a fixed dispersion of $V_{\rm cir}/\sqrt{2}
=140\kms$ for the Milky Way and $170\kms$ for the M31.
The assumption applies well in the halo, but not in the disk.

\subsection{Tidal stripping}

The outer region of a satellite will be stripped once it comes within
the strong tidal field of the host galaxy.  For simplicity, we assume
a rigid mass radius relation for the satellite, hence 
the tidal radius and the total mass of the satellite
are directly related by eq.(2), and 
\beq
{GM_s(r_t) \over r_t^3} = \left|\partial^2_R \phi(R)\right| = {V^2_{\rm cir} \over R^2},
\eeq
where $\phi(R)$ is the host potential at the radius $R$ (eq. 1).

\section{Results}

Fig.2 shows the decay of orbit of the hypothetical dwarf satellite
from an initial apocenter radius of 50 kpc, the norminal distance for
typical satellites.  Note this is also the apocenter of the
Sagittarius dwarf and its central cluster M54.  
Other Milky Way satellites, e.g., LMC, Draco and
Fornax, are beyond 50 kpc.  G1 and NGC1023-13 are also about 50 kpc
from their host galaxies.  We have taken the dashed line in Fig. 1 as
our satellite model (i.e. with $p=1$), but the result is largely the
same for the $p=2$ model.  We use the potential model SIS200 for the
Milky Way, but the conclusion does not change qualitatively when we
use the SIS250 potential model for M31.

We launch the orbits with different angular momentum.  A very low
angular momentum orbit (Fig.2a) could reach the current position of Omega Cen
at 6 kpc in one epicycle, but it bounces back to the original large orbit
because the specific orbital energy has not decayed by very much in
the brief period of one epicycle.  And the satellite is immediately
stripped into a globular cluster mass object at the peri-center, and
this greatly slows down the orbital decay because $dV/dt \propto M_s$.
The result is close to a system like G1 and NGC1023-13, which spends
most of its time relatively far from their host.  On the other hand a
high angular momentum orbit (Fig.2c) can hover at large radius, and
remain massive, but the dynamical decay is slow again because $dV/dt
\propto \rho(R) \propto R^{-2}$.  Even with intermediate angular
momentum (Fig.2b) the decay is very modest.  In summary Fig.2 shows
that we fail to decay the orbit of massive satellite from 50 kpc 
(the distance of G1) to a present apocenter of 6 kpc for Omega Cen.

What if we launch a satellite from 15 kpc just outside the edge of the
Milky Way disk?  To model the orbit of such closer satellite, a more
detailed model for the disk of the host galaxy is required.  because
the dynamical decay is enhanced by the high density of disk stars, and
the tidal effect is enhanced by the disk shocking.  Here we use
realistic potential models for MW and for M31 as given in Klypin,
Zhao, Somerville (2001).  These consist of three components, an NFW
halo, an exponential disk, and a bulge with a power-law nucleus.  We
ensure that a flat rotation curve, and the disk and halo density at
the solar neighbourhood are roughly produced.  Fig.3 shows how such orbit
could decay fairly efficiently, and after a full Hubble time the final
system is stripped off to a globular cluster-like object with an
orbital apocenter between 6-8 kpc.  The result is close to the orbit
of Omega Cen, but it would be somewhat larger given that we have only
a fraction of a Hubble time to decay the orbit.

The above simulations imply that the progenitor of Omega Cen might be
born within the radius of the Galactic disk, but on an inclined
retrograding orbit.  Any gas in the progenitor must have interacted
strongly with the direct rotating Galactic disk gas, and these interactions
might have left imprints in the stellar populations of Omega Cen.
Full exploration of the tide, ram pressure and the orbital parameter
space is beyond this contribution to the proceeding.

\section{Conclusion and discussion}

We have explored orbital decay for a dwarf satellite in a range of
galactic potentials.  Our calculations show that it is relatively easy
to strip off a dwarf galaxy to form a G1 like system on a large orbit.
But the progenitors of inner halo globular clusters can not be born at
a distance of 50 kpc from its host galaxy.  In particular we show that
there is not enough time to evolve a G1-like object, a transitional
object between a globular cluster and a dwarf galaxy, from a radius of
50 kpc to produce a Omega Cen like objects with an orbit of apocenter
of 6 kpc.  The progenitors must form relatively close to the Milky Way
disk and merge in the central few kpc.  The resulting dense globular
cluster would survive the merger and with only modest evolution from
the original orbit of the progenitor.  The question of the formation
of Omega Cen and its metal-rich stellar population remains open even
if Omega Cen was formed in a dwarf galaxy near the edge of the disk.
Could the metal rich gas survive the ram pressure stripping by the
Galactic disk gas?  We intend to continue the investigation of this
conundrum in forthcoming works (Fall, Gnedin, Livio, Meylan, Pringle,
Zhao 2002).

\acknowledgments 

I thank Ken Freeman, Oleg Gnedin, Mike Irwin, George Meylan, Jim
Pringle, Scott Tremaine, Floor van Leeuwen 
for enlightening discussions during the Omega Centauri conference at IoA,
and Mike Irwin for a careful reading of the draft.

\begin{figure*}
\includegraphics{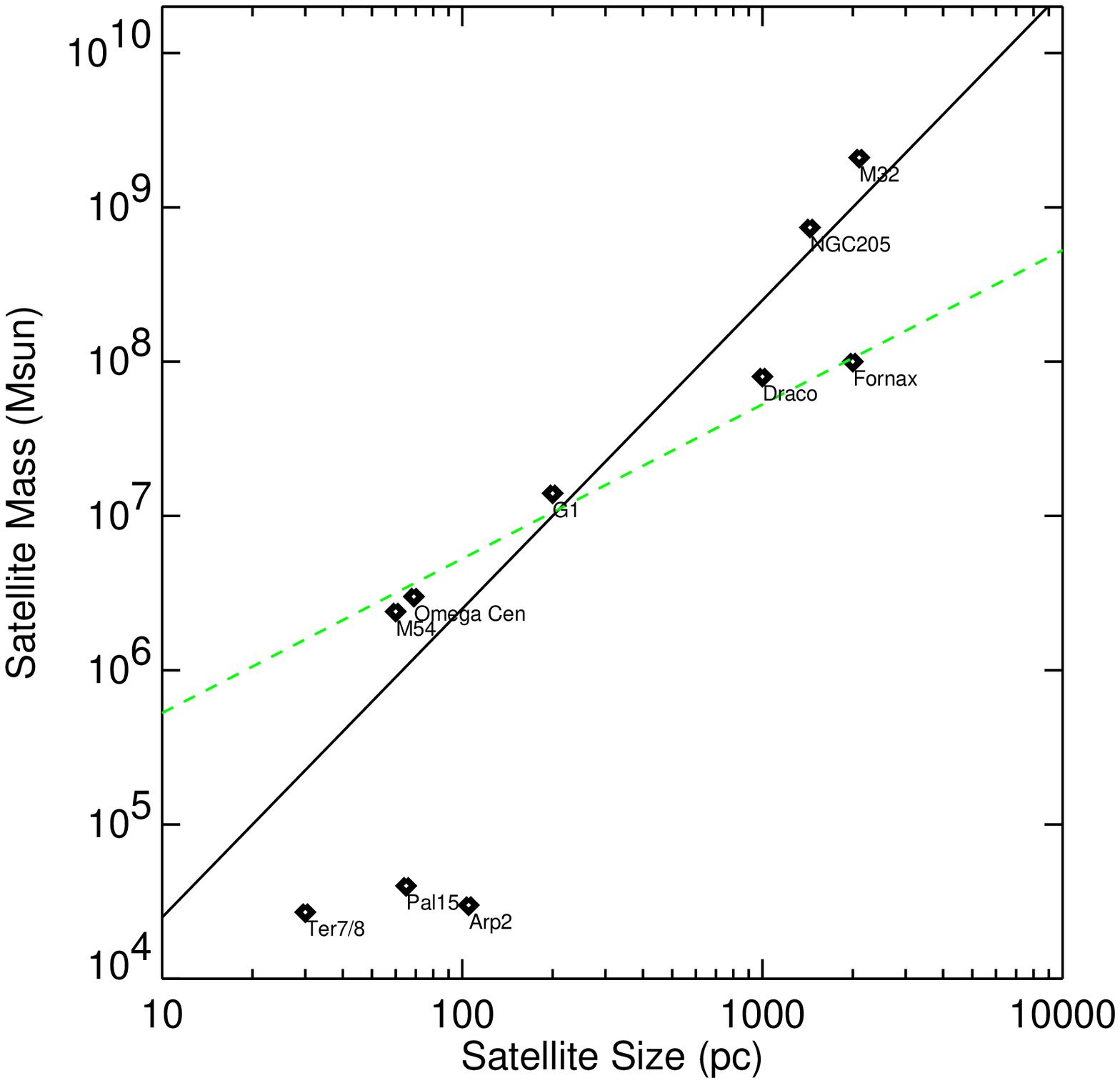}
\vspace{18cm}
\caption{
The tidal radius vs mass relation for 
the G1 cluster of M31 galaxy, 
Omega Cen of the Milky Way, the Draco and Fornax dSphs, 
the M32 and N205 dEs, and the globular clusters
of Sagittarius dSph (M54, Arp 2, Ter7/8, Pal 15).
Two lines are drawn to show our adopted 
mass-radius relations for a dwarf satellite in our simulation,
the dashed line with an isothermal power law, and the solid line with
a NFW cusp.  Data are taken from Harris (1996), Meylan et al. (2001),
Irwin \& Hatzidimitriou (1995), Cepa \& Beckman (1988).
}
\end{figure*}

\vfill\eject
\begin{figure*}
\includegraphics{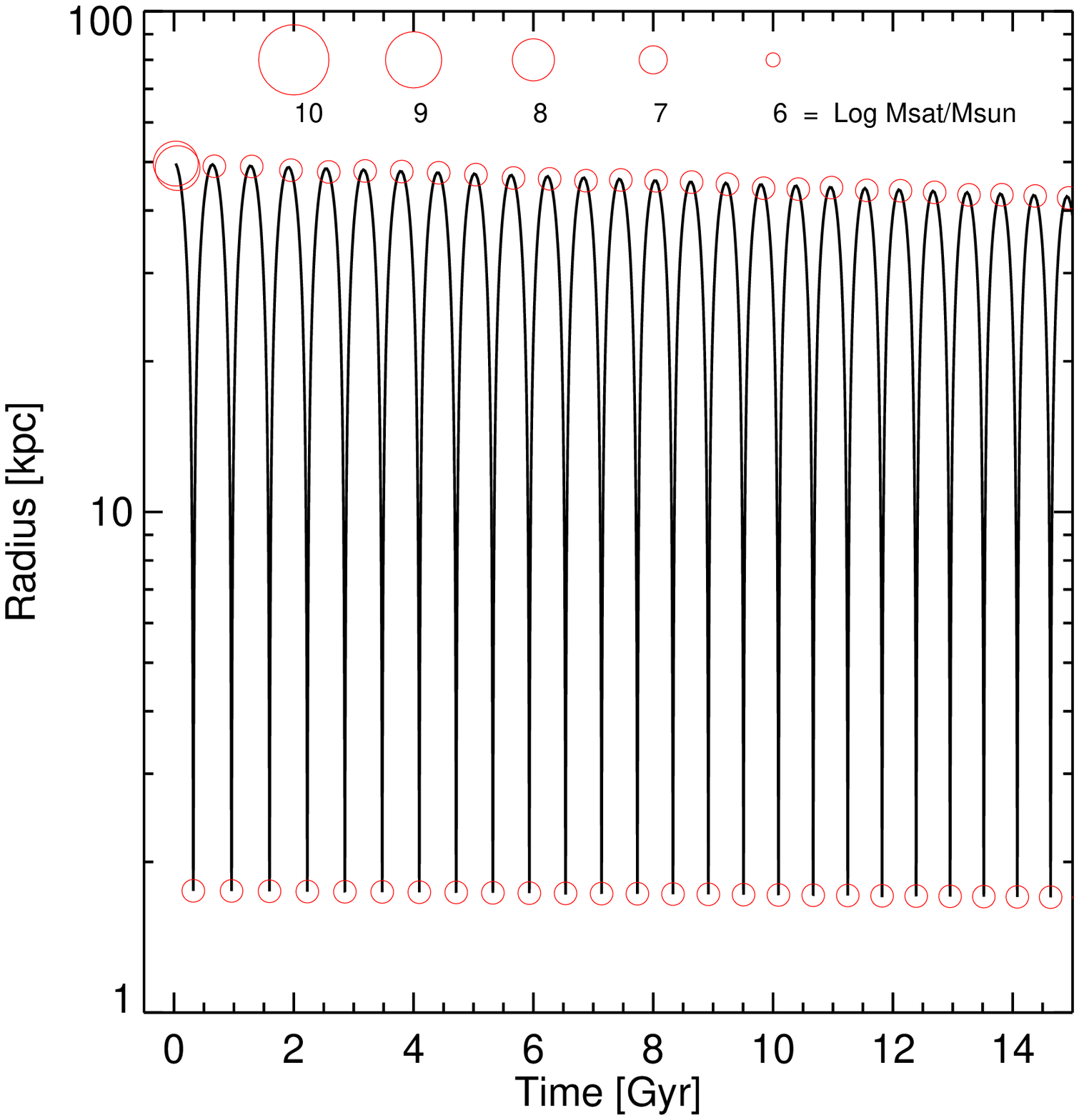}
\includegraphics{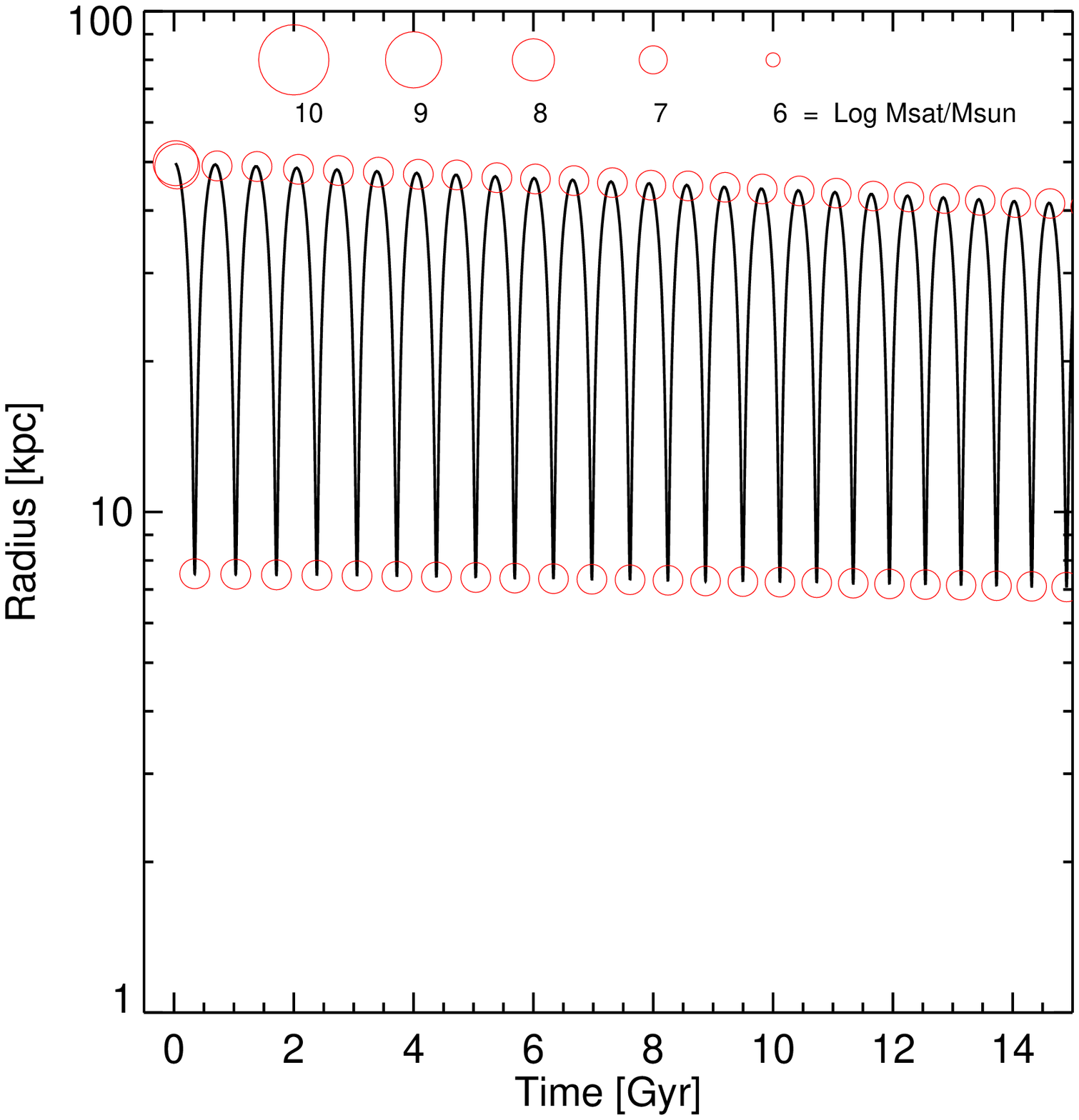}
\includegraphics{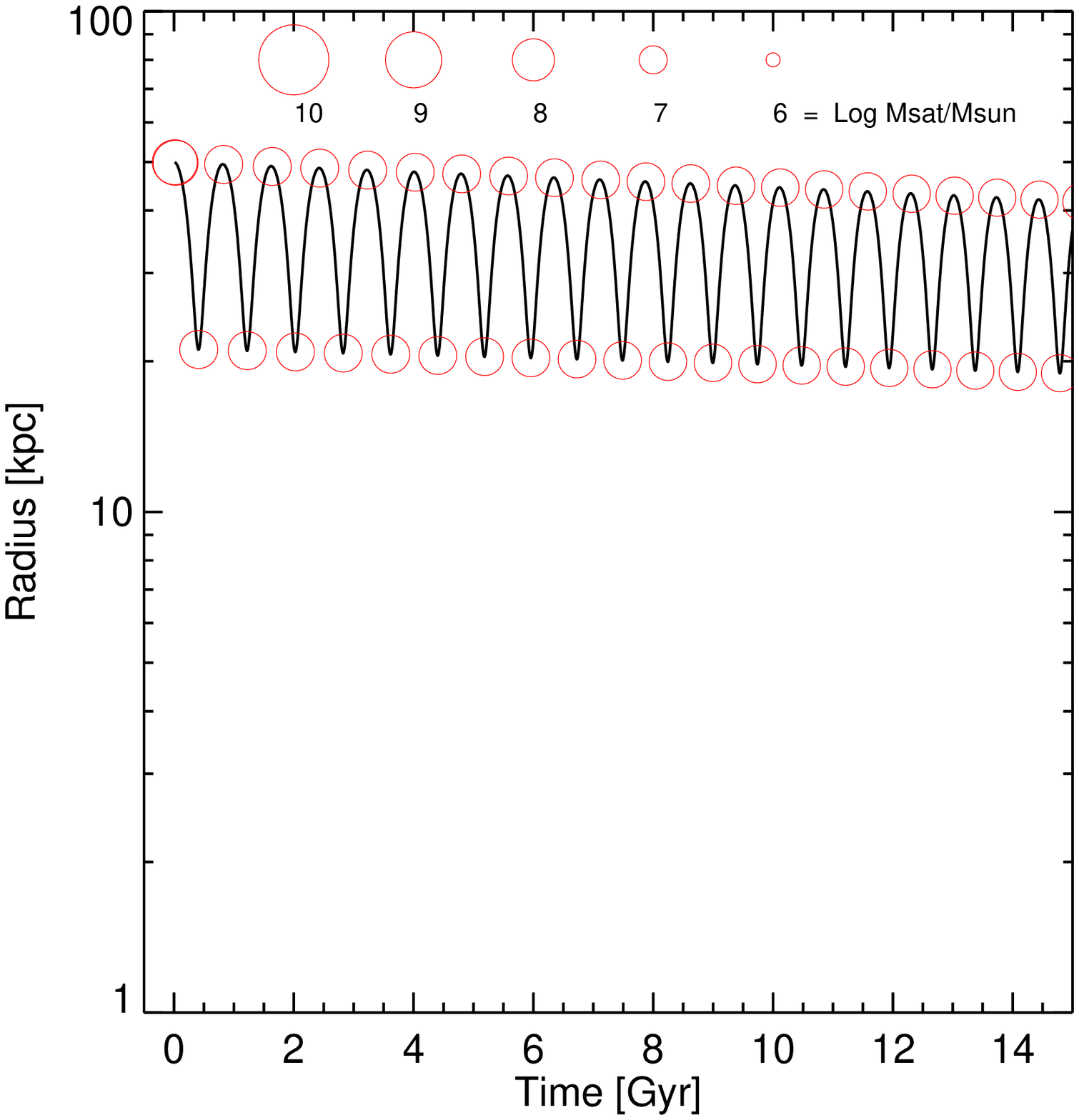}
\vspace{21cm}
\caption{
Simulated decay and stripping of a dwarf satellite with a $p=1$ mass profile
in a SIS potential of Vcir=200km/s.
Each panel shows the orbital radius (kpc) of the satellite
as a function of time (Gyr).  Sizes of red ellipsies indicate masses of
satellites at high/low tides.  The end result is 
generally a M54 or G1 like object with Log(Msat/Msun) $\sim$ 6 on a 
50 kpc orbit.
}
\end{figure*}

\vfill\eject
\begin{figure*}
\includegraphics{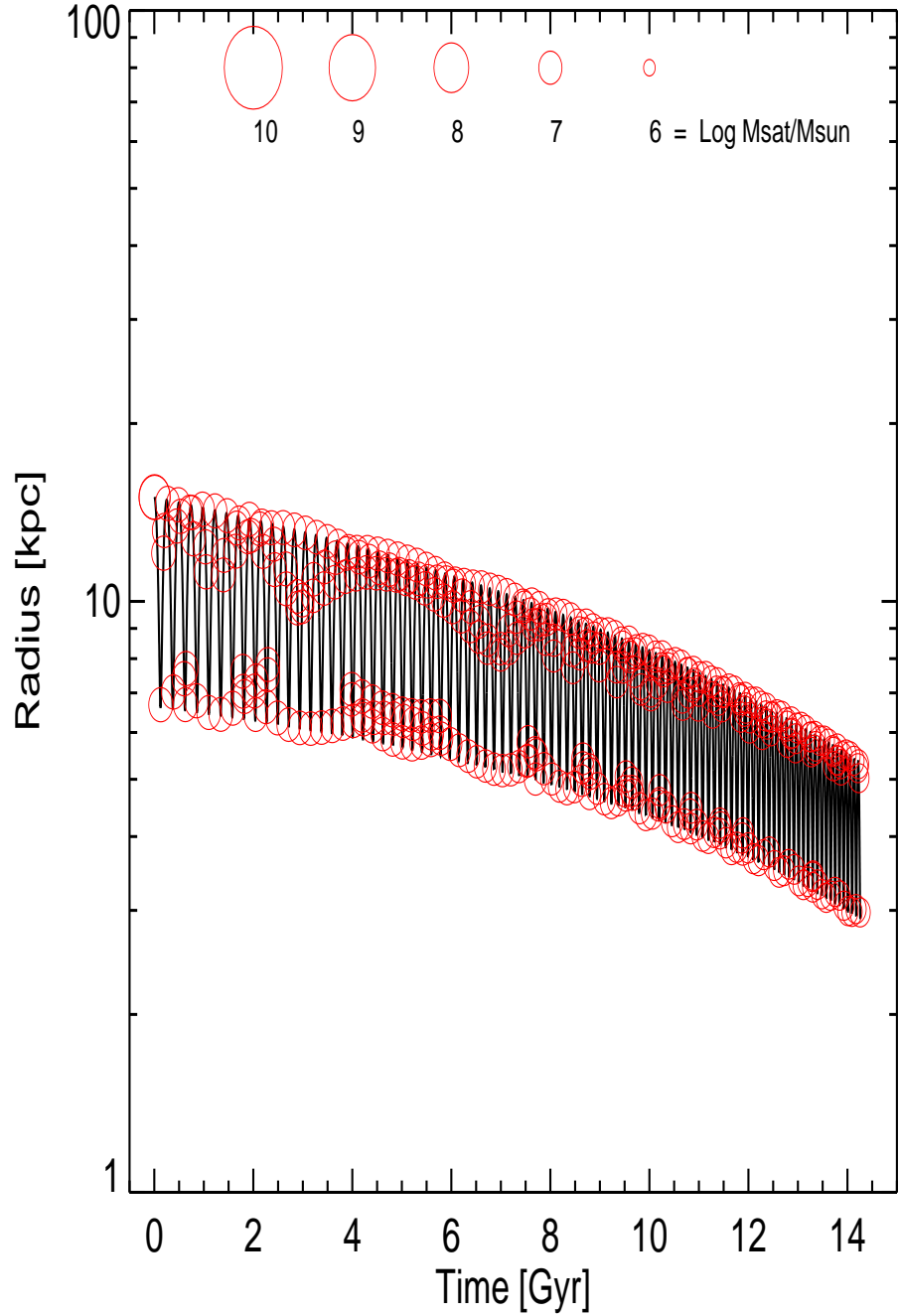}
\vspace{21cm}
\caption{
Similar to previous figure, but here for a realistic Milky Way potential
including a disk.  The satellite is launched from the edge of 
the disk at $R=15$ kpc and $|Z|=1$ kpc off the mid-plane with a
retrograding velocity.  Strong tides happen at disk-crossing.
After a Hubble time we end up with a system of similar mass 
($3\times 10^6\Msun$) and orbit ($1 \kpc \le R \le 6\kpc$) as Omega Cen.
}
\end{figure*}
\vfill\eject

\end{document}